\documentclass[reprint,nofootinbib,superscriptaddress,amsmath,amssymb,aps]{revtex4-1}
\usepackage{graphicx}
\usepackage{rotating}
\usepackage{dcolumn}
\usepackage{bm}
\usepackage{color}
\usepackage{mathptmx, textcomp}
\usepackage[latin1]{inputenc}
\usepackage{braket}
\usepackage[colorlinks,linkcolor=magenta,anchorcolor=cyan,citecolor=blue]{hyperref}
\usepackage[multiple]{footmisc}
\newcommand{\fig}[1]{Fig.\ref{#1}}
\def\be{\begin{equation}}
\def\ee{\end{equation}}
\def\ba{\begin{eqnarray}}
\def\ea{\end{eqnarray}}

\def\nn{\nonumber}

\newcommand{\eq}[1]{(\ref{#1})}

\def\q{\theta} \def\w{\omega}     \def\p {\pi}   \def\d {\delta} \def\f {\phi}     \def\l {\lambda} \def\z {\zeta} \def\x {\xi} \def\c {\chi} \def\b {\beta}  \def\m {\mu} \def\pd {\partial}\def\p {\pi} \def \inf {\infty}  \def \e { \varepsilon}
      \def\S {\Sigma}  \def\F {\Phi}      \def\.{\cdot}
\def\math {\mathcal}
\begin{document}
\title{Gedanken experiments at high-order approximation: nearly extremal Reissner-Nordstr\"{o}m black holes cannot be overcharged}
\author{Xin-Yang Wang}
\email{xinyang\_wang@foxmail.com}
\author{Jie Jiang}
\email{Corresponding author. jiejiang@mail.bnu.edu.cn}
\affiliation{Department of Physics, Beijing Normal University, Beijing, 100875, China}

\date{\today}

\begin{abstract}
The new version of the gedanken experiment proposed by Sorce and Wald has been used to examine the weak cosmic censorship conjecture (WCCC) for black holes at the second-order approximation of the matter fields perturbation. However, only considering the perturbation until the second-order approximation is incomplete because there is an optimal option such that the existing condition of the event horizon vanishes at second-order. For this circumstance, we cannot judge whether the WCCC is satisfied at this order. In our investigation, the $k$th-order perturbation inequality is generally derived. Using the inequalities, we examine the WCCC for nearly extremal Reissner-Nordst\"{o}m black holes at higher-order approximation. It is shown that the WCCC cannot be violated yet after the perturbation. From this result, it can be indicated that the WCCC is strictly satisfied at the perturbation level for nearly extremal RN black holes.
\end{abstract}

\maketitle
\section{Introduction}
The existence of black holes has been predicted by general relativity. For most black holes, there is a gravitational singularity at the center. Normally, the singularity should be surrounded by the event horizon and hides inside the black hole. If the event horizon vanishes, the naked singularity will be exposed to spacetime. Its emergence can block the well-define of spacetime and destroy the law of causality because of the curvature diverges at the position of the singularity. To avoid this situation, Penrose \cite{Penrose:1969pc} proposed the weak cosmic censorship conjecture (WCCC), which states that the singularity must be hidden inside the back hole and the observer at infinity cannot receive any information about the singularity.

To examine the validity of the WCCC, Wald \cite{Wald94} first proposed a gedanken experiment and demonstrated that the extremal Kerr-Newman (KN) black hole cannot be overcharged or overspun by dropping a test particle. Since then, utilizing this method, the WCCC for other black holes has been examined \cite{Cohen:1979zzb, Needham:1980fb, Semiz:1990fm, Bekenstein:1994nx, Semiz:2005gs}. However, the method has an inherent defect because the interaction between the particle and the background spacetime has been neglected. Moreover, Hubeny \cite{Hubeny:1998ga} found that if choosing a special particle with charge, the Reissner-Nordst\"{o}m (RN) black holes can be destroyed. In order to solve the defects, Sorce and Wald \cite{Sorce:2017dst} proposed a new version of the gedanken experiment to destroy the nearly extremal KN black holes at the second-order approximation of the perturbation that comes from the matter fields. The result showed that after the perturbation, the WCCC for KN black holes is still satisfied. Furthermore, using this method, the WCCC for other kinds of black holes is demonstrated to be valid \cite{An:2017phb, Ge:2017vun, Jiang:2019ige, Jiang:2019vww, Wang:2019bml, He:2019mqy, Jiang:2019soz, Jiang:2020btc, Jiang:2020mws}.

For the new version of the gedanken experiment, the first- and second-order perturbation inequalities are derived based on the Noether charge method proposed by Iyer and Wald \cite{Iyer:1994ys}. These inequalities reflect the null energy condition of the perturbation matter fields at first- and second-order approximations. After imposing the second-order perturbation inequality and the optimal condition of the first-order perturbation inequality, the result shows that the existing condition of the event horizon $h(\l)=M(\l)^2-Q(\l)^2-J(\l)^2/M(\l)^2$ under the second-order approximation can reduce to{
\ba\begin{aligned}
h(\l)\geq \left(\frac{(J^2-M^4)Q\d Q-2J M^2\d J}{M(M^4+J^2)}\l+M \e\right)^2\geq 0\,,
\end{aligned}\ea}
where $\e=r_h/M-1$ is a small parameter. This result shows the WCCC cannot be violated under the second-order approximation. However, there also exists an optimal option where the first two order perturbation inequalities are saturated and
\ba\begin{aligned}
\d Q=\frac{M^2[(M^4+J^2)\e-2J \d J\l]}{(M^4-J^2)Q\l}\,,
\end{aligned}\ea
which makes $h(\l)\simeq 0$ under the second-order approximation. When this happens, the sign of $h(\l)$ cannot be decided at this order. Therefore, we have to consider the higher-order approximation to examine the WCCC. For KN black holes, since the integral symplectic current is quite complex, to simplify the calculation and make the result more clear, we will use RN black holes to investigate whether the WCCC is still satisfied at the higher-order approximation.

In 1973, Boulware \cite{Boulware:1973tlq} investigated the process that a thin matter shell with energy and charge collapses to form RN black holes. It is shown that the shell can form a naked singularity when the energy density of the shell is negative. When it is positive, the singularity will be surrounded by the event horizon, and the WCCC cannot be violated. In this method, the distribution of the matter is just regarded as a discrete thin shell. However, in our universe, the distribution of matter is generally continuous. If we wish to examine the WCCC for RN black holes more general and closer to the real physical process, the matter fields should be regarded as continuously distributed outside the black hole. Fortunately, the continuous matter fields are taken into account in the new gedanken experiment. Therefore, in our research, we will use the new method as well to examine the WCCC for RN black holes under the perturbation that comes from the matter fields.

The organization of the paper is as follows. In Sec. \ref{sec2}, we discuss the spacetime geometry of RN black holes under the matter fields perturbation. In Sec. \ref{sec3}, we examine the WCCC for nearly extremal RN black holes under the fourth-order approximation. Furthermore, we attempt to discuss whether the WCCC is still satisfied under the $k$th-order approximation. In Sec. \ref{sec4}, some discussions and conclusions are given. In Appendix A, we introduce the definition of the Noether charge in Einstein gravity. In Appendix B, the first fourth-order perturbation inequalities are derived, and the general form of the $k$th-order perturbation inequality is further obtained.

\section{Perturbed geometry of RN black holes}\label{sec2}
For the four-dimensional Einstein-Maxwell gravitational theory, the Lagrangian is given as
\begin{equation}
	\boldsymbol{L} = \frac{1}{16 \pi} \left(R- F_{ab} F^{ab} \right) \boldsymbol{\epsilon}+\bm{L}_\text{mt}\,,
\end{equation}
where $\boldsymbol{F} = d \boldsymbol{A}$ is the strength of the electromagnetic field, $\boldsymbol{A}$ is the gauge potential of the electromagnetic field, $R$ is the Ricci scalar, { and $\bm{L}_\text{mt}$ is the Lagrangian of the extra matter fields. In the following, we will denote $\f$ to the collection of $g_{ab}$, $\bm{A}$ and the extra matter fields. When the extra matter fields vanish,} a class of static spherically symmetric solutions describing RN spacetimes in Eddington-Finkelstein coordinates is given as
\begin{equation}
	\begin{split}
		& ds^2 = - f(r) dv^2 + 2 dv dr + r^2 \left(d\theta^2 + \sin^2 \theta d\varphi^2 \right), \\
		& \boldsymbol{F} = \frac{Q}{r^2} dr \wedge dt.
	\end{split}
\end{equation}
In the line element of the spacetime, the expression of the blackening factor $f(r)$ is
\begin{equation}\label{bfpfr}
	f(r) = 1- \frac{2M}{r} + \frac{Q^2}{r^2}.
\end{equation}
The parameters $M$ and $Q$ in the blackening factor are associated with the mass and electric charge of the black hole. The radius of the event horizon $r_h$ is the largest root of the equation $f(r)=0$, i.e.,
\ba\begin{aligned}
r_h=M+\sqrt{M^2-Q^2}\,.
\end{aligned}\ea
Utilizing the expression of $r_h$, the surface gravity, the area of the event horizon, and the electric potential can be further given as
\begin{equation}
	\kappa = \frac{f'(r_h)}{2}, \qquad A_H = 4 \pi r_h^2, \qquad \Phi_H = \frac{Q}{r_h}.
\end{equation}

Subsequently, we consider a one-parameter family $\f(\l)$ of the field configurations, in which $\f(0)$ is a RN black hole solution. Each field configuration in the family is a spherical solution of the Einstein-Maxwell gravity sourced by some spherical matter fields which carry the energy and the electric charge in a finite region of the spacetime. When the value of $\lambda$ is small enough, the process can be regarded as a perturbation. In this family, the line element of the spacetime can be generally written as
\begin{equation}\label{dsd}
	ds^2 = - f(v, r, \lambda) dv^2 + 2 \mu (v, r, \lambda) dv dr + r^2 \left(d \theta^2 + \sin^2 \theta d \varphi^2 \right).
\end{equation}
When $f(v, r, 0) = f (r)$ and $\mu (v, r, 0) = 1$, the line element can degenerate into the case of the background spacetime.

Following a similar setup of Sorce and Wald \cite{Sorce:2017dst}, we only pay attention to the case that the perturbation is vanishing on the bifurcation surface $B$ and satisfying the stability condition. The stability condition states at sufficiently late times (where the perturbation matter fields all pass through the event horizon), the spacetime geometry can also be described by the class of static spherically symmetric solutions of the RN spacetime. It means that at sufficiently late times, Eq. (\ref{bfpfr}) can also be used to describe the spacetime geometry, just the parameters $M$ and $Q$ in the line element are replaced to $M(\lambda)$ and $Q(\lambda)$, i.e.,
\begin{equation}\label{bfpfrlambda}
	ds^2 (\lambda) = -f(r, \lambda) dv^2 + 2 dv dr + r^2 \left(d \theta^2 + \sin^2\q d \varphi^2 \right)\,,
\end{equation}
with
\begin{equation}
	f(r, \lambda) = 1 - \frac{2M(\lambda)}{r} + \frac{Q(\lambda)^2 }{r^2}\,.
\end{equation}

\section{Gedanken experiments at higher-order approximation}\label{sec3}
From here, we first examine whether the WCCC for nearly extremal RN black holes is still satisfied under the fourth-order approximation of the matter fields perturbation. Based on the stability condition, the spacetime geometry at late times becomes a static state, which means that checking the WCCC for RN black holes is equivalent to checking whether the event horizon also exists at sufficiently late times. For RN black holes, the existing condition of the event horizon is $M^2 - Q^2 \ge 0$. We just need to check whether the condition is also satisfied, i.e., $M(\lambda)^2 - Q(\lambda)^2 \ge 0$. Therefore, we define a function
\begin{equation}
	h(\lambda) = M(\lambda)^2 - Q(\lambda)^2
\end{equation}
and check the sign of the function $h(\lambda)$ to examine the WCCC. Under the fourth-order approximation of the parameter $\lambda$, we can get
\begin{equation}
	\begin{split}
h(\lambda)& \simeq M^2 - Q^2 + 2 \left(M \delta M -Q \delta Q \right) \lambda\\
 &+ \left(\delta M^2 - \delta Q^2 + M \delta^2 M -Q \delta^2 Q \right) \lambda^2 \\
& + \left(\delta M \delta^2 M - \delta Q \delta^2 Q + \frac{1}{3} M \delta^3 M - \frac{1}{3} Q \delta^3 Q \right) \lambda^3 \\
& + \frac{1}{12} \left(3 \delta^2 M^2 - 3 \delta^2 Q^2 + 4 \delta M \delta^3 M\right.\\
&\left. - 4 \delta Q \delta^3 Q + M \delta^4 M - Q \delta^4 Q \right) \lambda^4.
	\end{split}
\end{equation}
In the above expression, we have defined
\ba\begin{aligned}
\d^k\c=\left.\frac{d^k \c}{d\l^k}\right|_{\l=0}
\end{aligned}\ea
for the quantity $\c$.
Moreover, for nearly extremal RN black holes, since the value of the parameter $M$ extremely approaches to the value of the parameter $Q$, a small parameter $\epsilon$ can be defined as
\begin{equation}\label{defepsilon}
	\epsilon = \sqrt{M^2-Q^2}.
\end{equation}
With a similar consideration of \cite{Sorce:2017dst}, this parameter is chosen as the same order of the parameter $\lambda$. In order to simplify the calculating process and make the result more clearer, the value of the parameter $M$ can be set as $M = 1$ without loss of generality. Then, we have $Q = \sqrt{1 - \epsilon^2}$.

If we only consider the first-order approximation, the function $h(\l)$ can be expressed as{
\ba
h(\l)\simeq 2(\d M-\d Q) \lambda \,,
\ea}
where we have used the fact that the parameter $\epsilon$ is the same order as the parameter $\lambda$. In Appendix B, the null energy condition of the matter fields at the first-order approximation has been derived, which gives the first-order perturbation inequality
\ba\begin{aligned}
\d M-\F_H\d Q\geq 0\,.
\end{aligned}\ea
It implies that the function $h(\lambda)$ under the first-order approximation can reduce to
\begin{equation}\label{fohlambda}
	h(\lambda) \geq 0.
\end{equation}
It is shown that the WCCC cannot be violated under the first-order approximation. For $h(\lambda) > 0$ at the first-order approximation, { the higher-order approximation can be largely ignored.} However, there is an optimal option $\d M=\F_H\d Q$ such that $h(\l)=0$ under the first-order approximation. In this situation, the higher-order corrections will mainly affect the sign of $h(\l)$. Therefore, we should further consider the second-order approximation of $h(\l)$.

\begin{figure}
\centering
\includegraphics[width=0.47\textwidth]{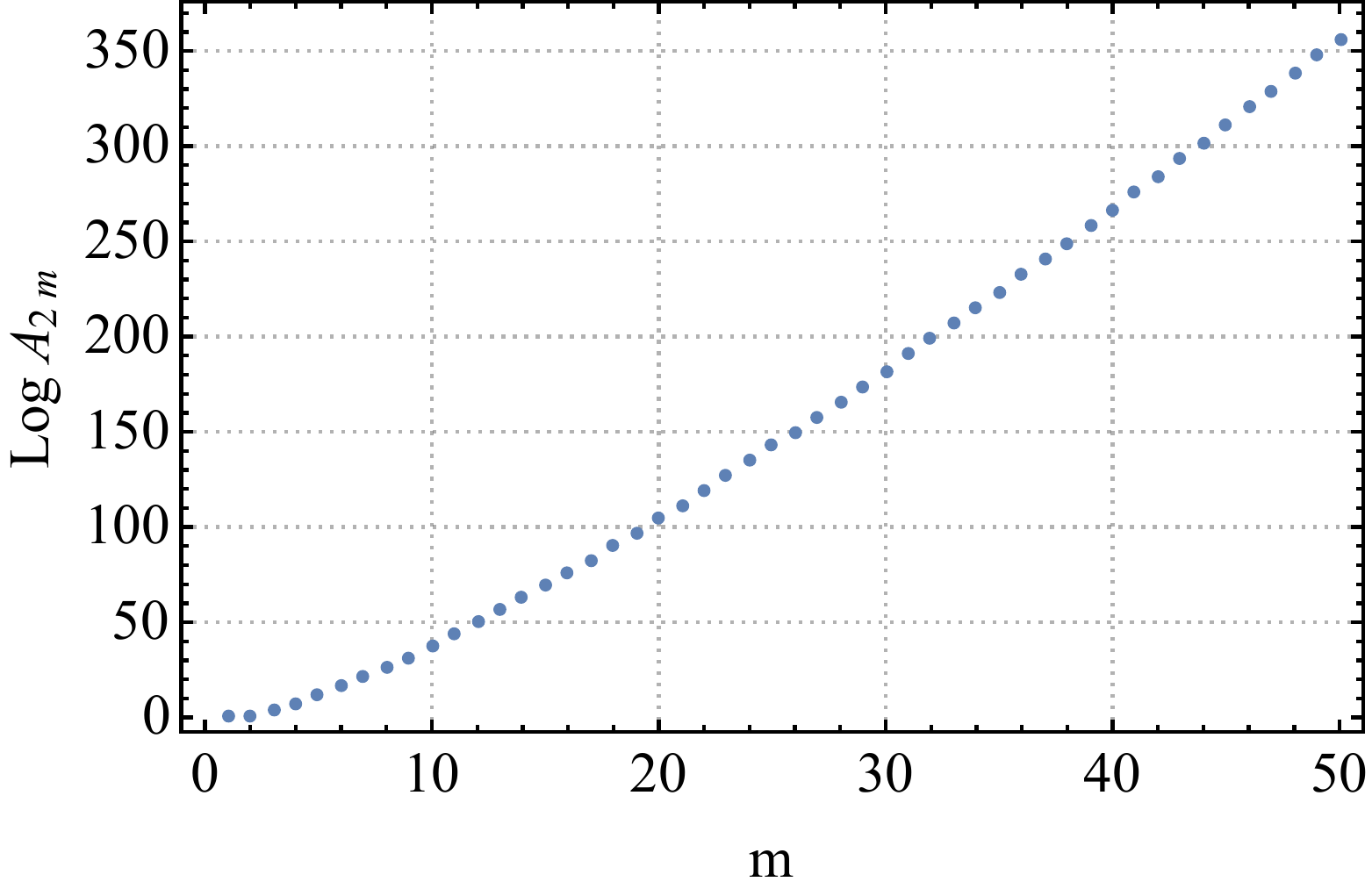}
\caption{Plot showing the value of the parameter $\log A_{2m}$.}\label{fig1}
\end{figure}

Under { the first-order optimal option}, the null energy condition of the matter fields at the second-order approximation gives the second-order perturbation inequality (see Appendix B)
\ba\begin{aligned}
\d^2M-\F_H \d^2Q-\frac{\d Q^2}{r_h}\geq 0\,.
\end{aligned}\ea
Together with the first-order optimal option, $h(\lambda)$ under the second-order approximation can reduce to{
\begin{equation}
	h(\lambda) \ge \left(\epsilon -  \lambda\delta Q \right)^2 \ge 0.
\end{equation}}
This result shows that the WCCC cannot be violated under the second-order approximation, which has also been obtained in \cite{Sorce:2017dst} for the nearly extremal KN black holes. However, analogous to the case of the first-order, there also exists a second-order optimal option, $\d^2M-\F_H \d^2Q-\d Q^2/r_h=0$ and $\d Q= \e/\l$, such that $h(\l)=0$ under the second-order approximation. In this situation, we should further consider the third-order approximation. When the first two order perturbation inequalities are both saturated, the null energy condition under the third-order approximation gives (see Appendix B)
\ba\begin{aligned}
\d^3M-\F_H \d^3 Q-\frac{3\d Q\d^2Q}{r_h}\geq 0\,.
\end{aligned}\ea
Together with the first two order optimal options, we have
\ba\begin{aligned}
h(\l)\geq 0
\end{aligned}\ea
under the third-order approximation. It is shown that the WCCC cannot be violated under the third-order approximation. However, under the third-order approximation, we can choose the optimal option as before, $\d^3M-\F_H \d^3 Q-3\d Q\d^2Q/r_h=0$. This optimal option makes $h(\l)=0$. In this case, the WCCC cannot be examined. Therefore we have to continue to consider the fourth-order approximation. Considering first three order saturation conditions, the null energy condition under the fourth-order approximation gives (see Appendix B)
\ba\begin{aligned}
\d^4M-\F_H \d^4 Q-\frac{3\d^2Q^2+4\d Q\d^3Q}{r_h}\geq 0\,.
\end{aligned}\ea
Therefore, utilizing the optimal options of the first three order perturbation and the fourth-order perturbation inequality, the function $h(\lambda)$ under the fourth-order approximation can be obtained as{
\begin{equation}
	h(\lambda) \ge \frac{1}{4} \left(\epsilon^2 - \l^2\delta^2 Q\right)^2 \ge 0.
\end{equation}}
From the result, we can clearly see that after the perturbation, the WCCC also cannot be violated under the fourth-order approximation of the perturbation. Obviously, there still exists an optimal option such that $h(\l)=0$ under the forth-order approximation. In principle, we should extend the discussion to any high-order approximation. The extension is straightforward but tedious. We attempt to extend this discussion into first $100$ order approximation and { summarize the results} as: for the $k$th-order approximation, we have
\ba\begin{aligned}
h(\l)&\geq 0\,,\quad\quad \quad\quad\quad
\text{for}\quad k=2(j+1)\\
h(\l)&\geq B_{k/2}(\l^{k/2}\d^{k/2} Q-A_{k/2}\e^{k/2})^2\geq 0\,,\text{for}\quad k=2j\,,
\end{aligned}\ea
under the first $(k-1)$ optimal options $\d^i Q=A_i(\e/\l)^i$ and $\d^l M-\frac{1}{2r_h}\sum_{i=0}^{l} C_k^i\d^{k-i}Q \d^i Q=0$ for any $i\leq [(k-1)/2]$ and $ l\leq k-1$, where $B_i$ is some positive parameter. When $i$ is an odd number greater than $1$, we have $A_i=0$. When $i$ is an even number, we show the value of $A_{2m}$ in \fig{fig1}. From the above results, it is not difficult to believe that the WCCC for nearly extremal RN black holes are always valid in any order approximation of the matter fields perturbation.

\section{Conclusions}\label{sec4}
In this research, we extended the new version of the gedanken experiment proposed by Sorce and Wald to the higher-order approximation of the perturbation that comes from the matter fields in the nearly extremal RN black holes. First of all, we generally derived the $k$th-order perturbation inequalities when the first $(k-1)$ order perturbation inequalities are saturated. Based on the general form of the perturbation inequality, we discussed the gedanken experiments up to $100$th order. The results show that the WCCC is always valid under the higher-order approximation. Therefore, we can infer that the WCCC is strictly satisfied at the perturbation level for the nearly extremal RN black holes.

\section*{Acknowledgement}
This research was supported by National Natural Science Foundation of China (NSFC) with Grants Nos. 11775022 and 11873044.

\section*{Appendix A: Noether charges in Einstein-Maxwell gravity}
In this appendix, we will review the Noether charge method proposed by Iyer and Wald \cite{Iyer:1994ys}. In Einstein gravity, the Lagrangian four-form is
\ba\begin{aligned}
\bm{L}=\frac{\bm{\epsilon}}{16\p}R\,.
\end{aligned}\ea
Considering the spacetime configuration with a one-parameter family, i.e., $g_{ab}(\l)$, we can { generally derive} the off-shell variation of the Lagrangian as
\begin{equation}\label{fdeltalag}
	\delta \boldsymbol{L} = \boldsymbol{E}_g^{ab} \delta g_{ab} + d \boldsymbol{\Theta}(g, \delta g)\,,
\end{equation}
where
\ba\begin{aligned}\label{egdeltag}
\boldsymbol{E}_g^{ab} &= -\frac{\boldsymbol{\epsilon}}{16\p} G^{ab}=-\frac{\boldsymbol{\epsilon}}{16\p}\left(R^{ab}-\frac{1}{2}Rg^{ab}\right)\,,\\
	\boldsymbol{\Theta}_{abc} (g, \delta g)& = \frac{1}{16 \pi} \bm{\epsilon}_{dabc} g^{de} g^{fg} \left(\nabla_g \delta g_{ef} - \nabla_e \delta g_{fg} \right).
\end{aligned}\ea
Here we have defined the $k$th order variation of the metric as
\ba\begin{aligned}
\d^k g_{ab}=\left.\frac{d^k g_{ab}(\l)}{d \l^k}\right|_{\l=0}\,.
\end{aligned}\ea
The symplectic current three form can be defined as
\begin{equation}\label{definitionomega}
	\boldsymbol{\omega} (g, \delta_1 g, \delta_2 g) = \delta_1 \boldsymbol{\Theta} (g, \delta_2 g) - \delta_2 \boldsymbol{\Theta} (g, \delta_1 g)\,,
\end{equation}
which can be explicitly written as
\begin{equation}
	\omega_{abc} = \frac{1}{16 \pi} \epsilon_{dabc} w^d,
\end{equation}
where
\begin{equation}
	w^a = P^{abcdef} \left(\delta_2 g_{bc} \nabla_d \delta_1 g_{ef} - \delta_1 g_{bc} \nabla_2 g_{ef} \right),
\end{equation}
with
\ba\begin{aligned}	
P^{abcdef} &= g^{ae} g^{fb} g^{cd} - \frac{1}{2} g^{ad} g^{be} g^{fc} - \frac{1}{2} g^{ab} g^{cd} g^{ef}\\
& - \frac{1}{2} g^{bc} g^{ae} g^{fd} + \frac{1}{2} g^{bc} g^{ad} g^{ef}.
\end{aligned}\ea
We set $\zeta^a$ as a infinitesimal generator of a diffeomorphism. Furthermore, replacing $\delta$ to $\mathcal{L}_\zeta$ in Eq. (\ref{fdeltalag}), one can define the Noether current three-form $\boldsymbol{J}_\zeta$ associated with $\zeta^a$, which can be written as
\begin{equation}\label{fjexp}
	\boldsymbol{J}_\zeta = \boldsymbol{\Theta} (g, \math{L}_\z g) - \zeta \cdot \boldsymbol{L}.
\end{equation}

On the other hand, it has been shown in Ref. \cite{Wald94} that the Noether current can also be formally written as
\begin{equation}\label{sjexp}
	\boldsymbol{J}_\zeta = \boldsymbol{C}_\zeta + d \boldsymbol{Q}_\zeta,
\end{equation}
in which{
\ba\begin{aligned}\label{chargegr}
	\left(\bm{Q}_\zeta \right)_{ab} &= - \frac{1}{16 \pi} \bm{\epsilon}_{abcd} \nabla^c \zeta^d\,,\\
	C_\zeta = \zeta \cdot C&\quad \text{with}\quad\bm{C}_{abcd} = \frac{1}{8\p}\bm{\epsilon}_{ebcd}G_a{}^e
\end{aligned}\ea}
are the Noether charge two form and the constrains of the Einstein gravity, respectively. Utilizing the above expressions, we can further obtain the first-order variational identities as
\ba\begin{aligned}\label{variation1}
&d\left[\delta \boldsymbol{Q}_\z - \z \cdot \boldsymbol{\Theta} \left(g, \delta g \right)  \right]\\
&= \boldsymbol{\omega} \left(g, \delta g, \mathcal{L}_\z g \right) - \z \cdot \boldsymbol{E}_g^{ab} \delta g_{ab} - \delta \boldsymbol{C}_\z.
\end{aligned}\ea

\section*{Appendix B: Derivation of the perturbation inequality}

In this appendix, we will give the detailed derivation process of each order perturbation inequalities which are involved in the paper. In the following, we consider the perturbation that comes from the collision process with a one-parameter family for the RN black holes as introduced in Sec. \ref{sec2}. First of all, we introduce a hypersurface $\Sigma = \S_0 \cup \Sigma_1$, in which $\S_0$ is a portion of the future event horizon of the background spacetime (i.e., the hypersurface $r=r_h$) starting from the bifurcation surface $B$ and continuing to { a cross section $B_1$ at a sufficiently late time}, and $\S_1$ is starting from the cross section $B_1$ and along the time-slice to go to the infinity. Note that in this choice, the hypersurface $\S$ is independent on the variational parameter $\l$ (i.e., $r_h$ is only the radius of the event horizon of the background spacetime and independent on $\l$). According to the stability condition, the spacetime geometry on the hypersurface $\Sigma_1$ can be described by the line element in Eq. \eq{bfpfrlambda}.

After replacing $\z^a$ by $\x^a=(\pd/\pd v)^a$, the integral of Eq. \eq{variation1} on the hypersurface $\Sigma$ gives
\ba\begin{aligned}\label{expression1}
&\int_{S_\inf}  \left[\frac{d\bm{Q}_\xi(\l)}{d\l} - \xi \cdot \boldsymbol{\Theta} \left(g(\l), \frac{dg(\l)}{d\l} \right) \right]  \\
&+\int_{\Sigma_1} \xi \cdot \boldsymbol{E}_g^{ab}(\l) \frac{dg_{ab}(\l)}{d\l} + \frac{d}{d\l}\left[\int_{\Sigma_1} \bm{C}_\xi(\l)\right]\\
&- \math{V}_0'(\l)-\math{E}_0(\l)= 0,
\end{aligned}\ea
where
{ \ba\begin{aligned}
\math{E}_0(\l)&=\int_{\S_0}\bm{\w}\left(g(\l), \frac{d g(\l)}{d\l}, \math{L}_\x g(\l)\right)\,,\\
\math{V}_0(\l)&= - \int_{\Sigma_0} \bm{C}_\xi(\l)\,.
\end{aligned}\ea
Here we have utilized the fact that $\xi^a$ is tangent to $\Sigma_0$ and therefore the integral of $\xi \cdot \bm{E}_g^{ab}$ is vanishing on $\Sigma_0$.} Meanwhile, we have used the assumption that the perturbation vanishes on the bifurcation surface $B$ and the stability condition. The stability condition makes $g_{ab}(\l)$ stable after the perturbation and $\math{L}_\x g_{ab}(\l)=0$ on the hypersurface $\S_1$. For the first term, performing  the explicit expression of the line element in Eq. \eq{bfpfrlambda}, we can obtain
\ba\begin{aligned}
\int_{S_\inf}  \left[\frac{d\bm{Q}_\xi(\l)}{d\l} - \xi \cdot \boldsymbol{\Theta} \left(g(\l), \frac{dg(\l)}{d\l} \right) \right]=M'(\l)\,.
\end{aligned}\ea
{ Utilizing the specific expression of the metric (\ref{bfpfrlambda}), we can easily check that the stress-energy tensor and the variation of the metric satisfy the following relation
\ba\begin{aligned}
T^{ab}(\l)\frac{d g_{ab}(\l)}{d\l}=0\,,
\end{aligned}\ea
where $T_{ab}(\l)=G_{ab}(\l)/8\p$ is the total stress-energy tensor of the electromagnetic field and perturbation matter fields. After using the above relation, the second term of Eq. \eq{expression1} vanishes.} For the third term, the straight calculation gives
\ba\begin{aligned}
\int_{\Sigma_1} \bm{C}_\xi(\l)=-\frac{Q^2(\l)}{2r_h}\,.
\end{aligned}\ea
For the { fourth term}, performing the explicit expression of the line element in Eq. \eq{dsd}, we can get
\ba\begin{aligned}
\math{V}_0(\l) & = - \frac{1}{8\p}\int_{\S_0}\bm{\epsilon}_{ebcd}(\l) \x^a T^e{}_a(\l) \\
& = \int_{\S_0}\tilde{\bm{\epsilon}} \m(v, r_h, \l)T_{ab}(\l)\x^a(dr)^b\,.
\end{aligned}\ea
Here we have denoted the volume element of the hypersurface $\S_0$ as $\tilde{\boldsymbol{\epsilon}} = dv \wedge \hat{\boldsymbol{\epsilon}}$, where $\hat{\boldsymbol{\epsilon}} = r^2 \sin \theta d\theta \wedge d\varphi $ is the volume element of the section of the event horizon. This term will reflect the null energy condition of the matter fields. To show this connection, we choose a null vector field
\begin{equation}
	l^a (\lambda) = \xi^a + \beta (\lambda) r^a,
\end{equation}
with
\begin{equation}
	r^a = \left(\frac{\partial}{\partial r} \right)^a, \qquad \beta (\lambda) = \frac{f(v, r_h, \lambda)}{2 \mu (v, r_h, \lambda)}
\end{equation}
on the hypersurface $\S_0$. Then, the null energy condition implies that
\ba\begin{aligned}
T_{ab}(\l)l^a(\l)l^b(\l)\geq 0\,.
\end{aligned}\ea
It can be proven that the null energy condition can be expanded as
\ba\begin{aligned}
&T_{ab} l^a (\lambda) l^b (\lambda) \\
&= \mu (v, r_h, \lambda) T_{ab} (\lambda) \xi^a (dr)^b + \beta (\lambda)^2 T_{ab} (\lambda) r^a r^b.
\end{aligned}\ea
Then, we have
\ba\begin{aligned}\label{null1}
\math{N}_0(\l)&=\math{V}_0(\l)+\tilde{\math{V}}_0(\l)\,,
\end{aligned}\ea
in which
\ba\begin{aligned}
\math{N}_0(\l)&=\int_{\S_0}\tilde{\bm{\epsilon}} T_{ab}(\l)l^a(\l)l^b(\l)\,,\\
\tilde{\math{V}}_0(\l)&=\int_{\S_0}\tilde{\bm{\epsilon}} \b^2(\l) T_{ab}(\l)r^a r^b\,.
\end{aligned}\ea
The null energy condition implies that $\math{N}_0(\l)\geq 0$. For the last term, we have
\ba\begin{aligned}
\math{E}_0(\l)=\frac{r_h}{2}\int_{v_0}^{v_1}dv\left[\pd_v p\pd_\l f-\pd_\l p\pd_v f\right]_{r=r_h}\,,
\end{aligned}\ea
where $v_0$ is the coordinate of the bifurcation surface $B$, $v_1$ is the coordinate of the hypersurface $B_1$, and we have denoted $p(v, r, \l)=1/\m(r, v, \l)$. { Combining} the above results, we can obtain the variational identity as
\ba\begin{aligned}\label{variationid}
M'(\l)-\frac{Q(\l)Q'(\l)}{r_h}=\math{V}_0'(\l)+\math{E}_0(\l)\,.
\end{aligned}\ea

\subsection*{The first-order perturbation inequality}\label{app1}

First, we consider the first-order perturbation inequality. After evaluating the value of the variation of Eq. \eq{variationid} on the background geometry (i.e., $\l=0$), we have
\ba\begin{aligned}\label{ineq1}
\d M-\F_H \d Q=\math{V}_0'(0)\,,
\end{aligned}\ea
where we have used the fact that the background geometry is static and therefore $\math{E}_0(0)=0$. For the null energy condition, according to Eq. \eq{null1}, we have
\ba\begin{aligned}
\math{N}_0(\l)\simeq \l \math{N}_0'(0)=\l \math{V}_0'(0)\geq 0
\end{aligned}\ea
under the first-order approximation of $\l$. Here we have used the fact $\b(0)=0$ for the background geometry such that $\tilde{\math{V}}_0'(0)=0$. The first-order variational identity in Eq. \eq{ineq1} reduces to
\ba\begin{aligned}
\d M-\F_H \d Q\geq 0\,.
\end{aligned}\ea
It is called the first-order perturbation inequality. This inequality is saturated when $\math{V}_0'(0)=0$. Using the explicit expression of the line element, this indicates that $\pd_v \d f(v, r_h)=0$.

\subsection*{The second-order perturbation inequality}\label{app2}
When the first-order perturbation inequality is saturated, the second-order perturbation inequality can be derived. Taking a variation of Eq. \eq{variationid} and evaluating it on the background geometry, we can further obtain
\ba\begin{aligned}\label{ineq2}
\d^2 M-\F_H \d^2 Q-\frac{\d Q^2}{r_h}=\math{V}_0''(0)+\math{E}_0'(0)\,.
\end{aligned}\ea
Using the first-order saturation condition $\math{V}'_0(0)=0$ and the fact that $\b(0)=0$ for the background geometry, the null energy condition under the second-order approximation gives
\ba\begin{aligned}
\math{N}_0(\l)\simeq\frac{\l^2}{2} \math{N}_0''(0)=\frac{\l^2}{2} \math{V}_0''(0)\geq 0\,.
\end{aligned}\ea
For the second term on the right hand side, we have
\ba\begin{aligned}
\math{E}_0'(0)&=\frac{r_h}{2}\int_{v_0}^{v_1}dv\left[\d f \pd_v \d p-\d p\pd_v \d f\right]_{r=r_h}\\
&=\frac{r_h}{2}\d f(v_1, r_h) \d p(v_1, r_h)=0\,,
\end{aligned}\ea
where we have used the stability condition such that $p(v_1, r_h, \l)=1$ in the last step. Summing the above results, the second-order variational identity \eq{ineq2} reduces to
\ba\begin{aligned}
\d^2 M-\F_H \d^2 Q-\frac{\d Q^2}{r_h}\geq 0\,.
\end{aligned}\ea
It is the second-order perturbation inequality. This inequality is saturated when $\math{V}''(0)=0$. Using the explicit expression of the line element and considering the first-order saturation condition, this indicates that $\pd_v \d^2 f(v, r_h)=0$.

\subsection*{The third-order perturbation inequality}\label{app3}
When the first two order inequalities are both saturated, the third-order perturbation inequality can be derived. Taking { two variations} of Eq. \eq{variationid} and evaluating it on the background, we have
\ba\begin{aligned}\label{ineq3}
\d^3 M-\F_H \d^3 Q-\frac{3\d Q \d^2 Q}{r_h}=\math{V}_0^{(3)}(0)+\math{E}_0''(0)\,.
\end{aligned}\ea
Using the first two order saturation conditions, the null energy condition under the third-order approximation implies that
\ba\begin{aligned}
\math{N}_0(\l)\simeq\frac{\l^3}{3!} \math{N}_0^{(3)}(0)=\frac{\l^3}{3!} \math{V}_0^{(3)}(0)\geq 0\,.
\end{aligned}\ea
Combining the two saturation conditions with the stability condition, the second term on the right hand side gives
\ba\begin{aligned}
\math{E}_0''(0)&=\frac{r_h}{2}\int_{v_0}^{v_1}dv\left[\d f \pd_v \d^2 p+2\d^2 f \pd_v \d p\right]_{r=r_h}\\
&=\frac{r_h}{2}\left[\d f \d^2 p+2\d^2 f \d p\right]_{v=v_1,r=r_h}\\
&=0\,.
\end{aligned}\ea
Summing the above results, the third-order variational identity \eq{ineq3} reduces to
\ba\begin{aligned}
\d^3 M-\F_H \d^3 Q-\frac{3\d Q \d^2 Q}{r_h}\geq 0\,.
\end{aligned}\ea
It is the third-order perturbation inequality. This inequality is saturated when $\math{V}^{(3)}(0)=0$. Using the explicit expression of the line element and together with the first two order saturation conditions, this indicates that $\pd_v \d^3 f(v, r_h)=0$.

\subsection*{The fourth-order perturbation inequality}\label{app4}
We will derive the fourth-order perturbation inequality when the first three inequalities are all saturated. Taking three variation of Eq. \eq{variationid} and evaluating it on the background geometry, we have
\ba\begin{aligned}\label{ineq4}
\d^3 M-\F_H \d^3 Q-\frac{3\d^2 Q^2+4\d Q \d^3 Q}{r_h}=\math{V}_0^{(4)}(0)+\math{E}_0^{(3)}(0)\,.
\end{aligned}\nn\\\ea
Analogous to the calculating process of the first three order inequalities, using the first three order saturation conditions, the null energy condition under the fourth-order approximation implies that $\math{V}_0^{(4)}(0)\geq 0$. Together with the stability condition, the second term on the right hand side gives
\ba\begin{aligned}
\math{E}_0''(0)&=\frac{r_h}{2}\int_{v_0}^{v_1}dv\left[\d f \pd_v \d^3 p+3\d^3 f \pd_v \d p+3\d^2 f \pd_v \d^2 p\right]_{r=r_h}\\
&=\frac{r_h}{2}\left[\d f \d^3 p+3\d^2 f \d^2 p+3\d^3 f \d p\right]_{v=v_1,r=r_h}\\
&=0\,.
\end{aligned}\ea
Summing the above results, the second-order variational identity \eq{ineq4} reduces to
\ba\begin{aligned}
\d^3 M-\F_H \d^3 Q-\frac{3\d^2 Q^2+4\d Q \d^3 Q}{r_h}\geq 0\,.
\end{aligned}\ea
It is the fourth-order perturbation inequality. This inequality is saturated when $\math{V}^{(4)}(0)=0$. Using the explicit expression of the line element and together with the first three order saturation conditions, this indicates that $\pd_v \d^4 f(v, r_h)=0$.
\\

\subsection*{The $k$th-order perturbation inequality}\label{appk}

Finally, we would like to perform the mathematical induction to prove the $(k+1)$th-order perturbation inequality when the first $k$ order perturbation inequalities are all saturated. We assume that the $k$th-order perturbation inequality is expressed as
\ba\begin{aligned}\label{ineqk}
&\d^k M-\frac{1}{2r_h}\sum_{i=0}^{k} C_k^i\d^{k-i}Q \d^i Q\geq 0\,,
\end{aligned}\ea
and the saturation of this inequality gives the additional condition $\math{V}^{(k)}(0)=0$ and $\pd_v\d^{k}f(v,r_h)=0$. Here $C_k^i$ is the binomial coefficient.  After that, we would like to prove that the expressions of the $(k+1)$th-order perturbation inequality and the related saturation condition are the same as the expressions of $k$th-order, just replacing the index $k$ as $k+1$. Taking $k$ variations on Eq. \eq{variationid} and evaluating it on the background, we have
\ba\begin{aligned}
&\d^{k+1} M-\frac{1}{2r_h}\sum_{i=0}^{k+1} C_{k+1}^i\d^{k+1-i}Q \d^i Q=\math{V}_0^{(k+1)}(0)+\math{E}_0^{(k)}(0)\,.
\end{aligned}\nn \ea
When the first $k$ order saturation conditions are taken into account, it is not difficult to verify that the null energy condition under the $(k+1)$th order approximation gives $\math{V}_0^{(k+1)}\geq 0$. For the second term of the right side, we have
\ba\begin{aligned}
\math{E}_0^{(k)}(0)&=\frac{r_h}{2}\sum_{i=0}^{i=k-1}C_k^i\int_{v_0}^{v_1}dv\left[\d^{i+1}f\pd_v \d^{k-i}p-\d^{i+1}p\pd_v \d^{k-i}f\right]_{r=r_h}\\
&=\frac{r_h}{2}\sum_{i=0}^{i=k-1}C_k^i\left[\d^{i+1}f\pd_v \d^{k-i}p\right]_{r=r_h,v=v_1}\\
&=0\,,
\end{aligned}\ea
where we have used the first $k$ order saturation inequality such that $\pd_v\d^i f=0$ for any $i\leq k$ and the stability condition $\d^j p=0$ for any $j$ at sufficiently late times. Summing the above results, the $(k+1)$th-order perturbation inequality reduces to
\ba\begin{aligned}
&\d^{k+1} M-\frac{1}{2r_h}\sum_{i=0}^{k+1} C_{k+1}^i\d^{k+1-i}Q \d^i Q\geq 0\,.
\end{aligned}\ea
This inequality is saturated when $\math{V}^{(k+1)}_0(0)=0$. Using the explicit expression of the line element and together with the first $k$ order saturated inequalities, it is not difficult to check that this condition implies that $\pd_v \d^{(k+1)} f(v, r_h)=0$. Until now, we have proved that the $k$th-order perturbation inequality and the related saturation condition can be expressed as Eq. \eq{ineqk}.

\end{document}